\documentclass[aps,pre,floatfix,nofootinbib,showpacs,twocolumn]{revtex4}
\usepackage{graphicx}
\usepackage{amsmath}
\usepackage{amssymb}
\newcommand{\be}{\begin{equation}}
\newcommand{\bea}{\begin{eqnarray}}
\newcommand{\bc}{\begin{center}}            
\newcommand{\ee}{\end{equation}}
\newcommand{\eea}{\end{eqnarray}}

\newcommand{\ec}{\end{center}}

\usepackage{default}
\newcommand{\bi}{\begin{itemize}}
\newcommand{\ei}{\end{itemize}}

\begin{document}
\author{George Thomas, Preety Aneja and 
Ramandeep S. Johal\footnote{electronic address:
rsjohal@iisermohali.ac.in}}
\affiliation{Indian Institute of Science Education and Research Mohali,\\
Sector 81, S.A.S. Nagar, P.O. Manauli, Mohali 140306, India}
\draft
\title{Informative priors and the analogy between quantum and
classical heat engines}
\begin{abstract}
When incomplete information about the control parameters is quantified as a prior distribution,
a subtle connection emerges between quantum heat engines and their classical analogs.
We study the quantum model where the uncertain parameters are the
intrinsic energy scales and compare with the classical models  where the intermediate temperature 
is the uncertain parameter. The prior distribution quantifying the  incomplete information has the 
form $\pi(x)\propto 1/x$ in both the quantum and the classical models. The expected
 efficiency calculated in near-equilibrium limit approaches the value of one third of Carnot efficiency.
\end{abstract}
\pacs{05.70.-a, 03.65.-w, 05.70.Ln, 02.50.Cw}
\maketitle
\section{Introduction}
Quantum heat engines (QHEs) are  novel tools to study the
 underlying thermodynamic properties  of quantum systems\cite{Johal2010, GRD2012, Scully, Lutz, ScullyPRL, Noah2010, GRD2011, Kieu2004}.
 The working substance in a QHE 
is  a few-level quantum system and can show interesting features 
 like quantum correlation, coherence and so on.
So QHEs may show unexpected behavior \cite{Scully, Lutz, ScullyPRL} which is not possible in the classical models of heat engines. 
But all these models
 are consistent with the 
 second law of thermodynamics. Recent studies \cite{Johal2010,GRD2012} showed that the expected behavior of
 certain  models of QHE exhibit classical thermodynamic features which points  out  interesting
 and novel connection between information and thermodynamics. In these models, the uncertain
 parameters are treated as in Bayesian approach.

In Bayesian approach to probability theory, prior distribution \cite{Jeffreys1939, Jaynes1968} or  known simply  as a prior,
 quantifies the prior knowledge 
about the uncertain parameter(s). Usually, there is some information available
even about the uncertain parameters, e.g. from the nature of the parameters or from
the physics of the problem. The prior which makes use of this 
knowledge is also addessed as an informative prior. So the right choice of the 
prior plays an important role in this approach.

In this paper, we discuss a quantum and a classical model of heat
 engine and estimate their performance.
In the quantum model we consider a pair of two level systems
 with energy level spacings $a_1$ and $a_2$. Reservoirs associated with the respective systems are at temperatures $T_1$ and $T_2$. These intrinsic energy level spacings can be controlled externally e.g. through external magnetic field. 
In the classical model, the pair of  two level systems is replaced by a
 pair of classical ideal gas systems.

 In the quantum model, the unknown parameters are the energy 
level spacings of the two level systems. But in  the classical case,
 the uncertain parameter is the intermediate temperature. To assign the prior, we
 invoke different observers who satisfy a consistency criterion and thus arrive at the  
prior for the unknown parameter. These prior distributions
 are used to estimate the expected behavior of
 thermodynamic quantities. Finally we compare
 the estimated values of the physical quantities obtained from
the quantum and the classical models. The main objective of this paper is 
to show the equivalence of the expected behavior of quantum and classical models 
under certain conditions.  Interestingly the expected efficiencies are also related to the  efficiencies
 at optimal performance 
for certain finite time models of Brownian heat engines.

The paper is organised as follows. In Section II,
we present the quantum model for heat engine and summarise its main features.
 In Section III, we discuss the prior chosen in the quantum model based on the initial
incomplete information about the internal energy scales of the working medium. In Section IV, we apply the
prior so
derived to estimate internal energies of the system. Further, we highlight a 
specific asymptotic limit in which the expected  behavior
becomes especially simple. Section V, we introduce the classical model with intermediate temperature as the uncertain
parameter.
 The final Section VI is devoted to conclusions and
future outlook.  
%
\section{The Quantum model}
\label{model}
Consider a pair of two-level systems labeled $R$ and $S$, with hamiltonians
$H_R$ and $H_S$
having energy eigenvalues $(0,a_1)$ and $(0,a_2)$, respectively. The hamiltonian
of the composite system is given by $H = H_R \otimes I + I\otimes H_S$. 
The initial state  is $\rho_{\rm ini} = \rho_R \otimes \rho_S$,  
where  $\rho_R$ and $\rho_S $ are thermal states corresponding to
 temperatures $T_1$ and $T_2$ ($< T_1$), respectively. Let ($r_1, r_2$) and 
($s_1, s_2$) be the occupation probabilities of each system, where 
\begin{equation}
r_1 = \frac{1}{(1+e^{-a_1/T_1})}, \qquad s_1 = \frac{1}{(1+e^{-a_2/T_2})}, 
\label{}
\end{equation}
with $r_2 = (1-r_1)$ and $s_2 = (1-s_1)$. We have set Boltzmann's constant
$k_{\rm B} = 1$.
The initial mean energy of each system is
\begin{equation}
 E_{\rm ini}^{(i)}=\frac{a_i}{(1+e^{a_i/T_i})},
\label{enin}
\end{equation}
where $i=1,2$ denote system $R$ and $S$ respectively.
Based on quantum thermodynamics \cite{Gyftopoulos, ABN2004,
AJM2008},
 the process of maximum work extraction
is identified as a quantum unitary process on the thermally isolated composite
system.
It preserves not just the magnitude of Von Neumann entropy of the composite
system,
but also all eigenvalues of its density matrix. It has been shown in these
works that for  $a_1 > a_2$, such a process minimises the
final energy if the final state is given by $\rho_{\rm fin} = \rho_S \otimes
\rho_R$.
Effectively, it means that in the final state the two systems \textit{swap} 
between themselves their initial probability distributions. The final energy of
each system at the end of work extracting transformation is
\begin{equation}
 E_{\rm fin}^{(i)}=\frac{a_i}{(1+e^{a_j/T_j})}.
\label{enfi}
\end{equation}
where $i \ne j$.
The average heat absorbed from system $R$ is 
defined as $Q_1 = {\rm Tr}[(\rho_{R}-\rho_{S})H_R]$, 
is given by
\begin{equation}
Q_1 = a_1  \left[ \frac{1}{\left( 1+e^{a_1/T_1}\right)} 
-  \frac{1}{\left( 1+e^{a_2 /T_2}\right)} \right].
\label{heat1}
\end{equation}
Similarly, the average heat released to system $S$ is 
defined as $Q_2 = {\rm Tr}[(\rho_S-\rho_{R})H_S]$, 
is given by
\begin{equation}
Q_2 = -a_2  \left[ \frac{1}{\left( 1+e^{a_1/T_1}\right)} 
-  \frac{1}{\left( 1+e^{a_2 /T_2}\right)} \right].
\label{heat2}
\end{equation}
The average work done in one cycle is $W=Q_1 + Q_2$.
To complete the cycle, the two systems are brought again in thermal contact 
with their respective reservoirs.
The operation of the machine as a  heat engine implies
${ W}\ge0$ and $Q_1 \ge0$, which is satisfied if
\begin{equation}
a_1 (T_2/T_1)  \le  {a_2} \le a_1.
\label{enginecriterion}
\end{equation} 
\section{Prior Distribution}
Now consider a situation in which the temperatures of the reservoirs are
given a priori such that $T_1 > T_2$, but the exact values of parameters $a_1$ and $a_2$
are uncertain. The prior information about these parameters may be summarised
as follows:
\begin{itemize}
\item $a_1$ and $a_2$ represent the same physical quantity, i.e. the level
spacing
for system $R$ and $S$ respectively, and so they can only take positive real values.
\item If the set-up of $R+S$ has to work as an engine, then criterion
in Eq. (\ref{enginecriterion}) must hold, whereby if one parameter is specified,
then the range of the other parameter is constrained.
\end{itemize}
Apart from the above conditions, we assume to have no information about $a_1$
and
$a_2$.
The question we address in the following is: What can we then infer about 
the expected behaviour 
of physical quantities for this heat engine ? 
We have suggested  a subjective or Bayesian approach  to address this question \cite{Johal2010,
GRD2012}. 
This implies that an uncertain parameter is assigned
 a prior distribution, which quantifies 
our preliminary expectation about the parameter to take a certain value. 
Thus the prior should be assigned by taking into account any prior
 information we possess about the parameters.
For example, if $a_1$ is specified, then the prior distribution
for $a_2$, $\pi(a_2|a_1)$ is 
conditioned on the specified value of $a_1$, and is defined in the range 
$[a_1 \theta, a_1]$, where $\theta = T_2/T_1$, because we know the set-up 
works like an engine if we implement Eq. (\ref{enginecriterion}).
We denote the prior distribution function for our problem by $\Pi(a_1, a_2)$.
To assign the prior, it seems convenient to involve two observers 
$A$ and $B$, who wish to assign priors for $a_1$ and $a_2$.
Based on the derivation given in \cite{GRD2012}, we find that
the prior for each parameter is 
\begin{equation}
 \Pi(a_i) =  \frac{1}{\ln{\left(\frac{a_{\rm max}}
{a_{\rm min}}\right)}}\frac{1}{a_i},
\label{aiprior}
\end{equation}
and the joint prior for the system acting as an engine, is given by
\begin{equation}
 \Pi(a_1, a_2) = 
\frac{1}{\ln\left({\frac{1}{\theta}}\right)\ln{\left(\frac{a_{\rm max}}
{a_{\rm min}}\right)}} \frac{1}{a_1 a_2}.
\label{jointpi}
\end{equation}
%
\section{Expected Values of Quantities}
In this section, we use the priors assigned above, to
find expected values for various physical
quantities related to the engine.
 The expected
 value of any physical quantity $X$ which may be function of $a_1$ and
$a_2$,  is defined as follows: 
\begin{equation}
\overline{X} = \int \int X\,\Pi(a_1,a_2)\,da_1 da_2.
\label{expectedvalue}
\end{equation}
 These expected 
values reflect the estimates by an observer who assigns the
priors. In principle, there are two ways to calculate the expected value
of some quantity which depends, in general, on the method used.
The  method by which the observer $A$ applies the joint prior is
based on the definition
\begin{equation}
\Pi(a_1, a_2) = \Pi(a_2|a_1)\Pi(a_1),  
\end{equation}
i.e. the prior for $a_1$ is assigned first, followed by $\Pi(a_2|a_1)$ which is the  conditional distribution 
of $a_2$ for a given value of $a_1$.
On the other hand, observer $B$ applies
\begin{equation}
\Pi(a_1, a_2) = \Pi(a_1|a_2)\Pi(a_2),  
\end{equation}
whereby the prior for $a_2$ is assigned first and $\Pi(a_1|a_2)$ represents the conditional distribution of $a_1$
for a given value of $a_2$.
%
\subsection{Internal energy}
We calculate the expected values of internal energies for 
 systems $R$ and $S$. These values can then be used to find the
expected work per cycle, heat exchanged and so on. 

(i) {\bf Initial state}: For a given $a_i$, the internal 
energy  $E_{\rm ini}^{(i)}$   is given by Eq. (\ref{enin}).
 The expected initial energy is defined as
\begin{equation}
\overline{E}_{\rm ini}^{(i)} = \int_{a_{\rm min}}^{a_{\rm max}}
 E_{\rm ini}^{(i)} \Pi(a_i) da_i,
\label{expectedin}
\end{equation}
where $i=1, 2$. Note that $E_{\rm ini}^{(i)}$ depends only on $a_i$, so we
 need to average over the prior for $a_i$ only. Using Eqs. (\ref{enin}) and
(\ref{aiprior}), we obtain 

\begin{eqnarray}
 \overline{E}_{\rm ini}^{(i)}&=&{\left[\ln{\left(\frac{a_{\rm max}}
{a_{\rm min}}\right)}\right]}^{-1}                                   \nonumber\\
&&\times \left[(a_{\rm max}-a_{\rm min})+T_i\ln{\left(\frac{1+e^{a_{\rm min}/T_i}}
{1+e^{a_{\rm max}/T_i}}\right)}\right]. \nonumber \\
\label{expein}
\end{eqnarray}   

(ii) {\bf Final state} : In this case, the internal energy of  $R$ as well as $S$,
 is function of both $a_1$ and $a_2$ (see (\ref{enfi})) and so the expected
values are
obtained by averaging over the joint prior, $\Pi(a_1,a_2)$. For instance, 
  the expected final energy of system $S$ (denoted by superscript (2)) as
calculated by $A$ is
\begin{eqnarray}
\overline{E}_{\rm fin}^{(2)}({A})&=&\int \int E_{\rm fin}^{(2)}\, \Pi(a_2|a_1)\Pi(a_1)\,da_1 da_2\nonumber\\
&=& K\int_{a_{\rm min}}^{a_{\rm max}}
\frac{1}{(1+e^{a_1/T_1})a_1}da_1\int_{a_1\theta}^{a_1}da_2,\nonumber\\
&=&K \left(1-\theta\right) \nonumber\\
&&\times\left[(a_{\rm max}-a_{\rm min})+T_1\ln{\left(\frac{1+e^{a_{\rm min}/T_1}}
{1+e^{a_{\rm max}/T_1}}\right)}\right].\nonumber \\
\label{EA}
\end{eqnarray}
where $K=[\ln{(1/\theta)}\ln{(a_{\rm max}/a_{\rm min})}]^{-1}$. Similarly as calculated by $B$, we have
\begin{eqnarray}
\overline{E}_{\rm fin}^{(2)}({B})&=&\int \int E_{\rm fin}^{(2)}\, \Pi(a_1|a_2)\Pi(a_2)\,da_1 da_2\nonumber\\
&=&K\int_{a_{\rm min}}^{a_{\rm max}}da_2
\int_{a_2}^{a_2/\theta}\frac{da_1}{(1+e^{a_1/T_1})a_1},
\label{efin2}
\end{eqnarray}
which cannot be solved analytically.

Now in general, the expected final energies of $S$, as given by Eqs. (\ref{EA}) and
(\ref{efin2})  
according to  $A$ and $B$, respectively, are not equal. 
One would expect that if the state of knowledge of $A$ and $B$ is similar,
then they should expect the same value for a given quantity. 
(Similar feature is also observed in the expressions for expected final energy of
system $R$.) 
\subsection{Asymptotic Limit}
\label{asymlim}
As remarked above, observers $A$ and $B$ should arrive at similar estimates
for physical quantities using their respective priors.
This happens in the limit, 
 $a_{\rm min}<<T_2$ and $a_{\rm max}>>T_1$. Then,
 Eq. (\ref{expein}) is approximated as
 \begin{equation}
\overline{E}^{(i)}_{\rm ini} \approx \frac{\ln 2}{\ln(\frac{a_{\rm max}}{a_{\rm
min}})}T_i.
\label{appei}
\end{equation}
The ratio $(a_{\rm max}/{a_{\rm min}})$ in the above may be large in magnitude, 
but is assumed to be finite.

Similarly, considering the final energies, it is remarkable to note that in this limit, not only the expected
energy
of a system ($R$ or $S$) calculated by either of the methods ($A$ or $B$), is the same 
but also its value for system $R$ or $S$ is also equal. 
 Thus from Eqs. (\ref{EA}) and
(\ref{efin2}), we have (omitting 
the observer index)
\begin{equation}
\overline{E}^{(i)}_{\rm fin} \approx
\frac{\ln 2}{\ln(\frac{a_{\rm max}}{a_{\rm min}})}\frac{(1-\theta)T_1}{\ln
(\frac{1}{\theta})}.
\label{appef}
\end{equation} 
%
 Further insight may be obtained if we 
 estimate the final temperatures of systems $R$ and $S$, after the  work
extraction process. 
Note that if values of both $a_1$ and $a_2$ are specified, the temperatures ($T_i'$)
of 
the two systems after work extraction, are given by \cite{AJM2008}
\begin{equation}
 T_1' = T_2 \frac{a_1}{a_2},\quad \mbox{and} \quad
T_2' = T_1 \frac{a_2}{a_1}.
\end{equation}
In general, the two final
temperatures are different from each other.
However following the subjective approach, when
we look at the expected values of the final temperatures as calculated by
$A$ or $B$, we find
\begin{equation}
 \overline{T}_1' = \overline{T}_2' = T_1 \frac{(1-\theta)}{\ln(1/\theta)}.
\label{tempfin}
\end{equation}
It is interesting to find that the assignment of the prior is such
that the two systems are expected to finally arrive at a common temperature. 
Going back to Eqs. (\ref{appei}) and (\ref{appef}) for the energies, 
we see that they satisfy a simple
relation $\overline{E}_{\rm ini}^{(i)} \propto T_i$
and $\overline{E}_{\rm fin}^{(i)} \propto \overline{T}_i'$.
This is analogous to the thermodynamic behavior of a classical ideal gas.

Next, the expected values of the heat exchanged between  system $i$  and the corresponding reservoir
is given by
 $\overline{Q}_i=\overline{E}_{\rm ini}^{(i)}-\overline{E}_{\rm fin}^{(i)}$. 
 $\overline{Q}_i>0$ ($\overline{Q}_i<0$) represents heat absorbed 
 (released) by the system. Then the expressions for the heat exchanged with the 
reservoirs in the said limit are as follows:
\begin{equation}
\overline{Q}_1 \approx \frac{\ln2}{\ln\left(\frac{a_{\rm
max}}{a_{\rm min}}\right)} \left(1+\frac{(1-\theta)}{\ln\theta} \right)T_1,
\label{q1asym}
\end{equation}
and
\begin{equation}
\overline{Q}_2 \approx \frac{\ln2}{\ln\left(\frac{a_{\rm
max}}{a_{\rm min}}\right)} \left(1+\frac{(1-\theta)}{\theta \ln\theta}
\right)T_2.
\label{q2asym}
\end{equation}
Now the  expected work per cycle is defined as: 
 $\overline{W}=\overline{Q}_1+\overline{Q}_2$.  
Thus the  efficiency may be defined as $\eta = 1+\overline{Q}_2/\overline{Q}_1$.
Explicitly, using Eqs. (\ref{q1asym}) and (\ref{q2asym}) we get
\begin{equation}
\eta = 1 + \frac{\theta \ln\theta +(1-\theta)}{\ln\theta +(1-\theta)}.
\label{effby3}
\end{equation}
This is the efficiency at which the engine is expected to operate
for a given $\theta$. The above value is function only of the ratio
 of the reservoir temperatures. 
 
 We note that the constant of proportionality in Eqs. (\ref{appei})
and (\ref{appef}),
which is ${\ln 2}\cdot\left({\ln({a_{\rm max}}/{a_{\rm min}})}\right)^{-1}$, 
can be related with heat capacity. The expected value of initial heat 
capacity of system $i$, defined as 
\begin{equation} 
\overline{C}_i = \int_{a_{\rm min}}^{a_{\rm max}}  C_i \Pi(a_i) d a_i,
\end{equation}
 where we know for a two level system, the canonical heat capacity at constant
volume is $ C_i=(a_i/T_i)^2(\exp{(a_i/T_i)})/(1+\exp{(a_i/T_i)})^2$.

In particular, for the asymptotic limit, the leading term yields  
\begin{equation}
 \overline{C}_i \equiv \overline{C} \approx\frac{\ln2}{\ln\left(\frac{a_{\rm
max}}{a_{\rm min}}\right)}.
\label{expcapacity}
\end{equation}
This  value is independent of temperature of the system and thus 
indicates an analogy with a constant heat capacity thermodynamic system.  

Thus the requirement of consistency
between the results of $A$ and $B$ implies, in an asymptotic limit,
that the behavior expected from minimal
prior information is the one
which shows simple thermodynamic features such as constant heat
capacity and equality of subsystem temperatures upon maximum
work extraction. 
%
\section{Classical Model}
\label{classicalmodel}
In this section, we discuss a classical model of the heat engine, within
the subjective approach. 
Consider two thermodynamic systems at initial temperatures $T_1$ and $T_2$ 
($< T_1$),
such that the heat capacity $C$ can be assumed to be constant i.e. 
the systems behave like classical ideal gases. 
 Further consider 
the maximum work extraction by coupling the two systems to a reversible
work source. This process preserves the total entropy, i.e. $\bigtriangleup S=0$.
 Let us assume that at some stage, the temperatures of the systems are
$T_a$ and $T_b$, respectively. Entropy conservation leads to 
\be
T_b = \frac{T_1 T_2}{T_a}.
\label{Tb}
\ee 
The above equation relates $T_b$ and $T_a$, implying that given
a value of one of them, the value of the other is fixed.
The work $W$ extracted from the system is given by the decrease in internal energy,
\be
W =  C(T_1 + T_2 - T_a - T_b).
\label{w}
\ee
Similarly, the heat absorbed from the initially hotter system will be
\be
Q_1 = C(T_1-T_a).
\ee
Now we consider the situation in which we have lack of information
about the exact values of these intermediate temperatures $T_a$
and $T_b$. We want to estimate the properties of this engine,
taking into account the prior information we have about the
parameters. Now it is clear that we have to assign the prior
either to $T_a$ or $T_b$, because the two parameters
are related by Eq. (\ref{Tb}). Imagine two observers
$A$ and $B$ who respectively take $T_a$ and $T_b$, as the uncertain parameter
for the considered thermodynamic process. Thus from
the perspective of $A$, the work is given by
 \be
W =  C\left(T_1 + T_2 - T_a - \frac{T_1 T_2}{T_a}\right),
\label{wa}
\ee
while from $B$'s point of view
\be
W =  C\left(T_1 + T_2 - T_b - \frac{T_1 T_2}{T_b}\right).
\label{wb}
\ee
Further we assume that $A$ and $B$ assign the same functional form 
for their priors and the range over which the parameters
can take values is also the same. Thus probability distribution $P$
for the  temperatures $T_a$ and $T_b$ is defined in the range $[T_1, T_2]$. 
Now due to the constraint relating $T_a$ and $T_b$, the probabilities
assigned to any pair of values related by Eq. (\ref{Tb}), should be the same. This
means
\be
P(T_a) dT_a = P(T_b) dT_b.
\label{P}
\ee
Using (\ref{Tb}) and (\ref{P}), we get:
\be
P(T_a) = \frac{1}{\ln(1/\theta)}\frac{1}{T_a},
\label{P1}
\ee
where $\theta = T_2/T_1$.
\subsection{Estimated Work and Efficiency}
The work estimated by $A$ is defined as:
\be
\langle W \rangle = \int_{T_2}^{T_1} W P(T_a) dT_a, 
\ee
where $W$ is given by Eq. (\ref{wa}). Thus 
\bea
\langle W \rangle &= & C T_1\left [(1 + \theta )  +
 \frac{2 (1 - \theta)}{\ln {\theta}} \right].
\eea
Similarly, the estimate for the heat absorbed is given by:
\be
\langle Q_1 \rangle = CT_1\left[1 + \frac{(1 - \theta)}{\ln {\theta}}\right].
\ee
The efficiency, $\overline{\eta} = \langle W \rangle / \langle Q_1 \rangle$, is given by:
\be
\overline{\eta} = 1 + \frac{ \theta \ln{\theta} + (1 - \theta)}{\ln {\theta} + (1 -\theta)}.
\label{eta3}
\ee
The observer $B$ also arrives at the same estimates for 
work and efficiency. 
Now the expected values of intermediate temperatures calculated by $A$ and $B$ after 
the work extraction process are
\begin{equation}
\langle T_a\rangle=\langle T_b\rangle= T_1\frac{(1-\theta)}{\ln{(1/\theta)}}.
\end{equation}
Moreover, all these estimates are also the same as derived from a quantum 
heat engine in the asymptotic limit in section \ref{asymlim}. 
%
\section{Conclusions}
We have analysed the case of heat engines by assuming 
uncertainty in the exact values of the internal energy scales
of the working medium. 
We have suggested the appropriate prior distributions for the uncertain parameters based 
on  prior information. In the case of quantum model, where the level spacings 
 are the unknown parameters, the expected values of work, heat and efficiency 
are equivalent to the expected values calculated from the classical
 model, where the intermediate temperature is the unknown parameter. 
Moreover in both the models, expected temperatures 
are equal at the end of work extraction.
So the quantum model with two unknown parameters shows similar behavior as the
 classical model, with a single uncertain parameter. In this point, it
is interesting to analyse a special case of the quantum model with a single unknown parameter. 
Such kind of situation arises when the efficiency ($\eta=1-a_2/a_1$) of the
engine is given. This  simplifies the problem to that of 
 a single uncertain parameter, either $a_1$ or $a_2$. Introducing two observers as
discussed in \cite{GRD2012}, we get the functional form of the prior as
 $\pi(a_i)\propto1/a_i$.
We can then calculate the expected work per cycle. In the asymptotic limit,
the expression for expected work reduces to 
\begin{equation}
 \overline{W}\approx \overline{C}\eta\left(T_1-\frac{T_2}{(1-\eta)}\right).
\label{wasym}
 \end{equation}
Let us compare the above expression with its classical analog.
The classical model  of the heat engine discussed in  section \ref{classicalmodel}
  has an efficiency $\eta=1-T_a/T_1=1-T_2/T_b$, when $T_a$ and $T_b$ are specified. Substituting this efficiency in
Eqs. (\ref{wa}) and (\ref{wb}), we get the  expression for work same  as the  quantum case  given in Eq. (\ref{wasym}).   
 
The efficiency at the maximum expected work (Eq.(\ref{wasym})) is  equal to $1-\sqrt{\theta}$, the well known Curzon-Ahlborn efficiency \cite{CA1975}. 
 In near equilibrium limit, this efficiency can be expanded as 
\begin{equation}
\eta^* \approx \frac{(1-\theta)}{2} + \frac{(1-\theta)^2}{8} + O(1-\theta)^3.                               
\label{expneta2}
\end{equation}
So this efficiency appears at the maximum value of the expected work in the asymptotic limit for the quantum model with one  uncertain parameter as well as at maximum work for 
the classical model without any uncertain parameter.
This efficiency falls in a certain universality class 
\cite{Tu2008,Esposito2010}, where the leading term in the near equilibrium expansion is half of Carnot efficiency($\eta_c$).
At this point, it is interesting to analyse the near equilibrium
 expansion of the efficiency expressed in Eqs. (\ref{effby3}) and (\ref{eta3}):
\begin{equation}
\eta^* \approx \frac{(1-\theta)}{3} + \frac{(1-\theta)^2}{9} + \frac{8(1-\theta)^3}{135}+ O[(1-\theta)^4].                              
\label{expneta3}
\end{equation}
This expected efficiency is obtained  in two cases, from the quantum model where two
 parameters ($a_1$ and $a_2$) are uncertain and also from  the classical model with a single uncertain 
parameter ($T_a$ or $T_b$). The leading term in the near equilibrium expansion of this efficiency 
is  $\eta_c/3$ instead of  $\eta_c/2$ observed in Eq. (\ref{expneta2}).  As another example, the efficiency at 
maximum power  for an irreversible Brownian heat engine  \cite{Zhang2006} when optimized 
over the load and barrier height is given by
\begin{equation}
\eta^* = \frac{2(1-\theta)^2 }{3-2\theta(1+\ln\theta)-\theta^2}. 
 \label{effzh}
\end{equation}
Expanding this efficiency for the near-equilibrium regime, we get
\begin{equation}
\eta^* \approx \frac{(1-\theta)}{3} + \frac{(1-\theta)^2}{9} + \frac{(1-\theta)^3}{18} + O[(1-\theta)^4].                              
\label{expneta4}
\end{equation}
\begin{figure}[ht]
\includegraphics[width=8cm]{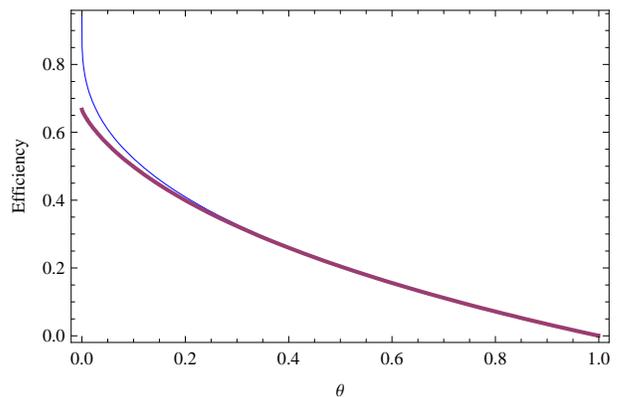}
\caption{The efficiency in Eq. (\ref{eta3}) and (\ref{effzh}) are plotted versus $\theta$. 
The lower curve (thick line)
is the efficiency at maximum power for the Brownian heat engine. The upper curve is the efficiency obtained from the classical model
with intermediate temperature as the single uncertain parameter. The same expected efficiency is also obtained in the asymptotic limit of the quantum model when two intrinsic energy scales of the working medium are uncertain.} 
\label{fig1}
\end{figure}
The similarity between Eqs. (\ref{expneta3}) and (\ref{expneta4}) up to the second order,
suggests there might be a new universality class which includes these efficiencies. 
In Fig. 1, the efficiencies given in (\ref{eta3}) and (\ref{effzh}) are plotted. 
To conclude, when all the unknown parameters are treated with prior probabilities, the expected behavior
 of quantum model with two internal energy scales  as the uncertain parameters is similar to the classical model with intermediate temperature as the single uncertain parameter. Interestingly, the expected efficiency in 
both cases, approaches $\eta_c/3$ value, in the near-equilibrium limit. 
Further, the analysis based on subjective treatment of the incomplete information, suggests 
a new line of enquiry of a possible connection with optimal performance of the finite time 
irreversible models of heat engines.  
\section{Acknowledgements}RSJ thanks the organizers of FQMT-2011 conference and in particular, 
Prof. Theo M. Nieuwenhuizen and Prof. Vaclav Spicka for the invitation to Prague. RSJ is supported by Department of 
Science and Technology, India under the project No. SR/S2/CMP-0047/2010(G). GT thanks IISER Mohali for research fellowship. PA is supported by UGC through Junior research fellowship.
%
 
%
%
\

\end{document}